\documentclass[11pt,twoside]{article}
\usepackage{asp2014}

\aspSuppressVolSlug
\resetcounters

\begin{document}

\title{Standardisation of Data Formats in Gamma-ray astronomy}

\author{C. Nigro$^1$ and T. Hassan$^2$ \newline 
on behalf of the \textit{Data Formats for Gamma-ray Astronomy} forum contributors}
\affil{$^1$Institut de Física d'Altes Energies (IFAE), Barcelona, Spain; \email{cosimo.nigro@ifae.es}}
\affil{$^2$Centro de Investigaciones Energéticas, Medioambientales y Tecnológicas (CIEMAT), Madrid, Spain; 
\email{tarek.hassan@ciemat.es}}

\begin{abstract}

The operation of the future Cherenkov Telescope Array (CTA), the next generation of Imaging Atmospheric Cherenkov Telescopes (IACTs), as an Open Observatory requires data products and analysis tools to be accessible and usable by the scientific community. This requirement has impelled gamma-ray astronomers to develop open-source science tools and standardised data formats. The objective of this presentation is to provide a perspective on the current effort to define a prototypical format for high-level gamma-ray astronomical data. The initiative, started by the IACT community in 2016, has gained in the years 2018 and 2019 full recognition thanks to the first public data releases in such format by  current-generation IACTs, followed by a series of papers employing it either to test the capabilities of open-source gamma-ray analysis tools or to showcase examples of reproducible, multi-instrument analysis.

\end{abstract}

\section{Introduction}
With the noticeable exception of \textit{Fermi}-LAT, current instruments for gamma-ray astronomy rely on independent proprietary software and data for their scientific analyses. These technical differences not only prevent multi-instrument analyses to be performed within a common framework, but also impede any legacy data products to be disseminated without releasing the corresponding analysis software. Furthermore, the next instrument for gamma-ray astronomy at Very High Energies (VHE, $E > 100\,{\rm GeV}$), CTA \citep{cta_science}, will be the first IACT to be operated as an Open Observatory. The requirement on its analysis tools and data to be accessible and usable to the astronomical community has led to the development of open-source science tools, \texttt{ctools} \citep{ctools} and \texttt{gammapy} \citep{gammapy}, and to the definition of a preliminary standardised data format for gamma-ray data \citep{gadf_2016}. This initiative, focus of our review, will be described in Section \ref{sec:gadf}. In Section \ref{sec:dl3_projects} we will instead describe the projects that implemented such data standard and used it either for the purpose of validating the above mentioned science tools or to illustrate the ease of multi-instrument analysis once data standardisation has been achieved.

\section{The \textit{Data Formats for Gamma-ray Astronomy} Forum}
\label{sec:gadf}

\begin{figure}
    \centering
    \includegraphics[scale=0.15]{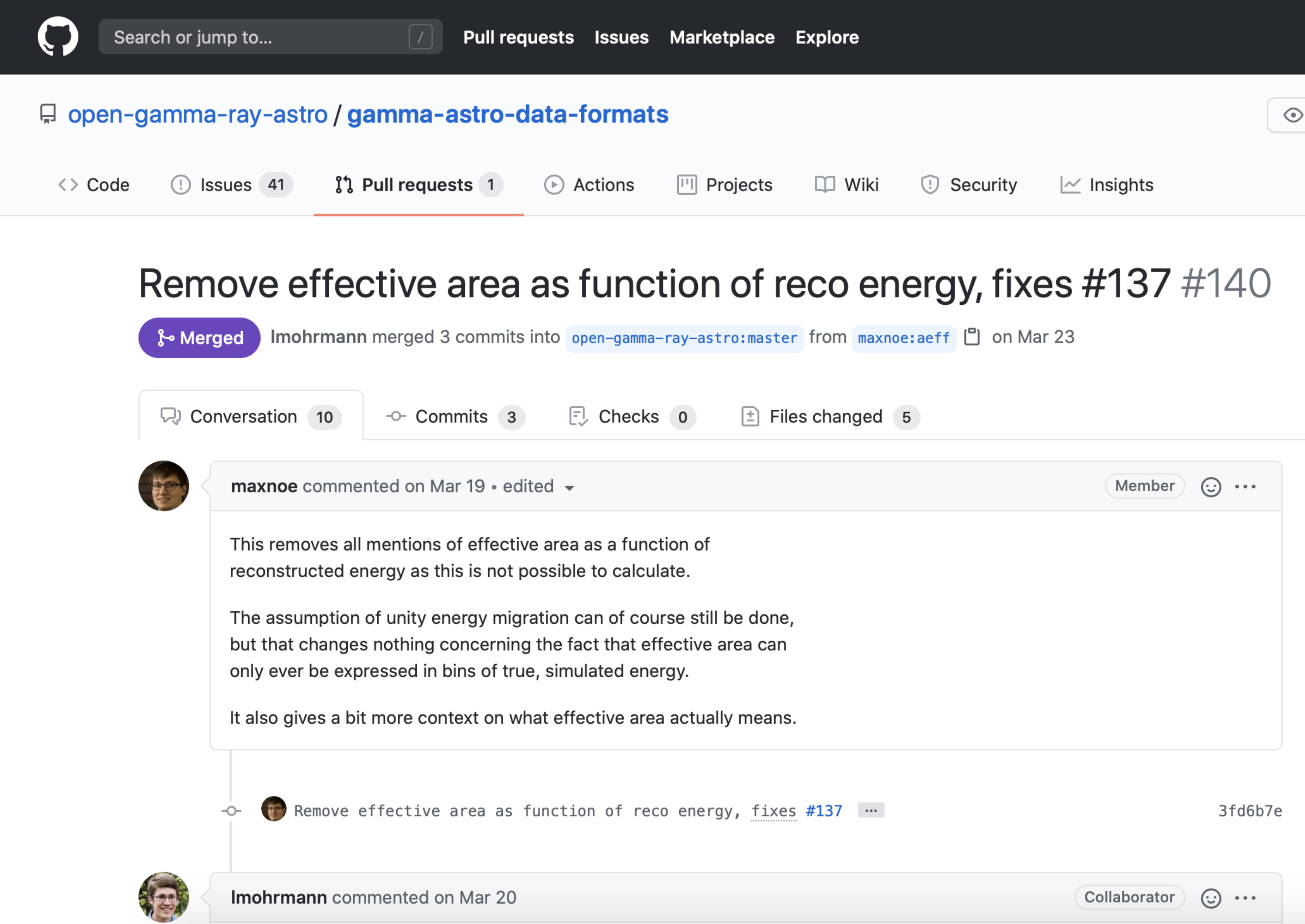}
    \includegraphics[scale=0.15]{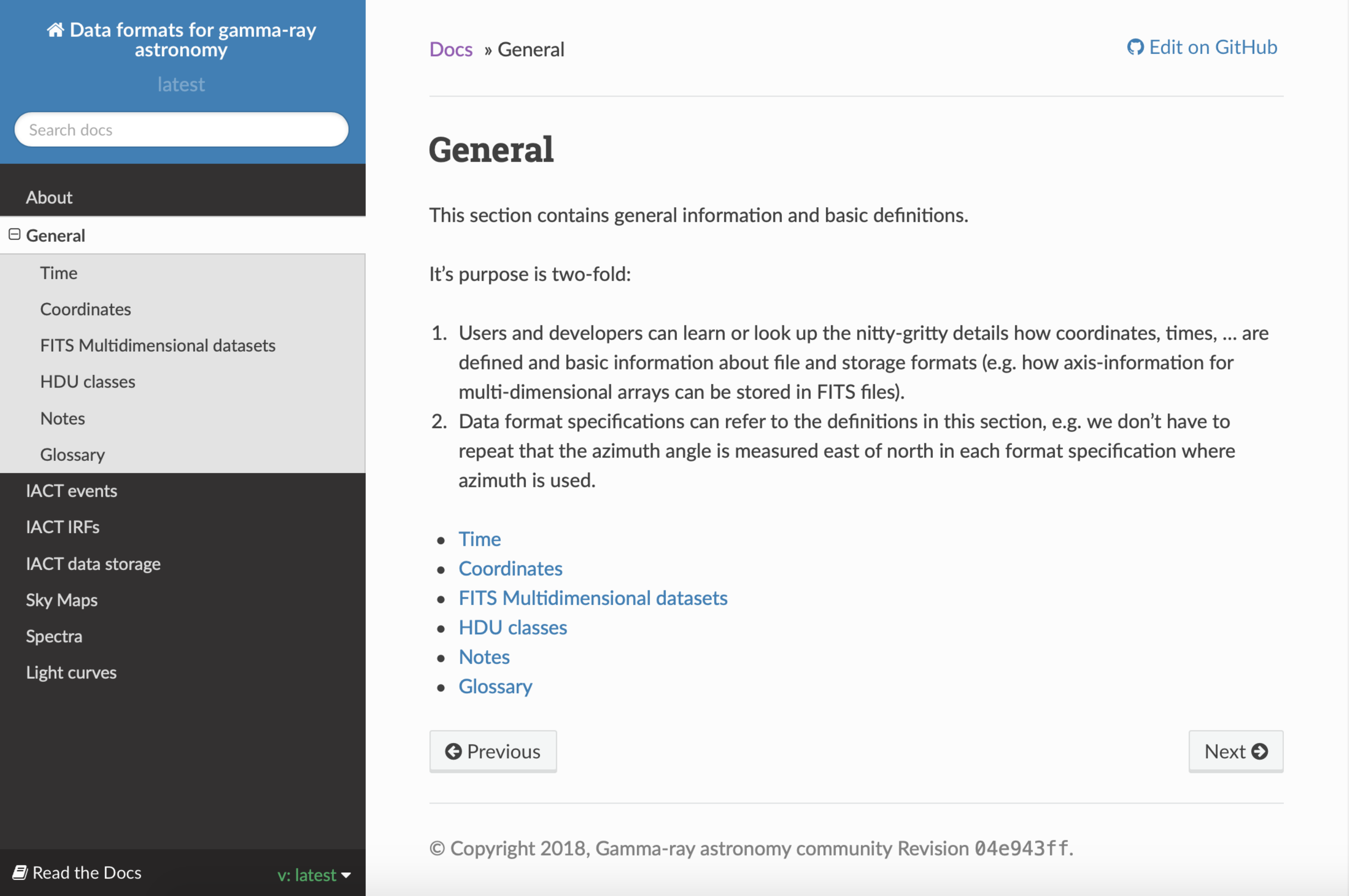}
    \caption{\texttt{github} repository (left) and online documentation (right) hosting the \textit{Data Formats for Gamma-ray Astronomy} specifications.}
    \label{fig:gadf}
\end{figure}

The discussion around a common format for high-level gamma-ray astronomical data was started in early 2016 and has been channeled in the so-called \textit{Data Formats for Gamma-ray Astronomy} forum (\citealt{gadf_2016}). 
The data level chosen to be standardised is the so-called Data Level 3 (DL3, \citealt{dl3}) which encapsulates lists of events classified as $\gamma$ rays and the Instrument Response Function (IRF) in the data units of a \texttt{FITS} file \citep{fits}.
The specifications of the format, establishing the content and the naming scheme of the \texttt{FITS} files are written in the form of a \texttt{sphinx} documentation hosted on \texttt{github}\footnote{\href{https://github.com/open-gamma-ray-astro/gamma-astro-data-formats}{https://github.com/open-gamma-ray-astro/gamma-astro-data-formats}} and built via \texttt{readthedocs}\footnote{\href{https://gamma-astro-data-formats.readthedocs.io/en/latest/}{https://gamma-astro-data-formats.readthedocs.io/en/latest/}}, as shown in Figure \ref{fig:gadf}. The open access to the format allows both gamma-ray astronomers producing the actual data and science tools developers to modify or expand it following the \texttt{github} workflow. 

\section{Projects using the Data Level 3 Format}
\label{sec:dl3_projects}
Though initiated with the objective of CTA, the DL3 format can accommodate the data of current-generation IACTs. In what follows we present some projects already implementing and testing DL3 data products and open-source science tools. 

\subsection{The H.E.S.S. First Public Test Data Release}
The H.E.S.S. collaboration was the first to make publicly available a small sample of its earliest observations (2004-2006) in DL3 format \citep{hess_dl3}\footnote{\href{https://www.mpi-hd.mpg.de/hfm/HESS/pages/dl3-dr1/}{https://www.mpi-hd.mpg.de/hfm/HESS/pages/dl3-dr1/}}. The release contains $30\,{\rm h}$ of observations of sources representing different science cases (Crab Nebula, PKS~2155-304, MSH~15-52 and RX~J1713.7-3946) and $20\,{\rm h}$ of observations of sky patches devoid of gamma-ray sources, usable to build a background model (off-source observations). The aim of this release is to advertise the endeavour on the data standardisation and to allow for a double feedback: both on the side of the data format and on the science tools that will employ it. 

\subsection{Towards open and reproducible multi-instrument analysis in gamma-ray astronomy}

\begin{figure}
    \centering
    \includegraphics[scale=0.26]{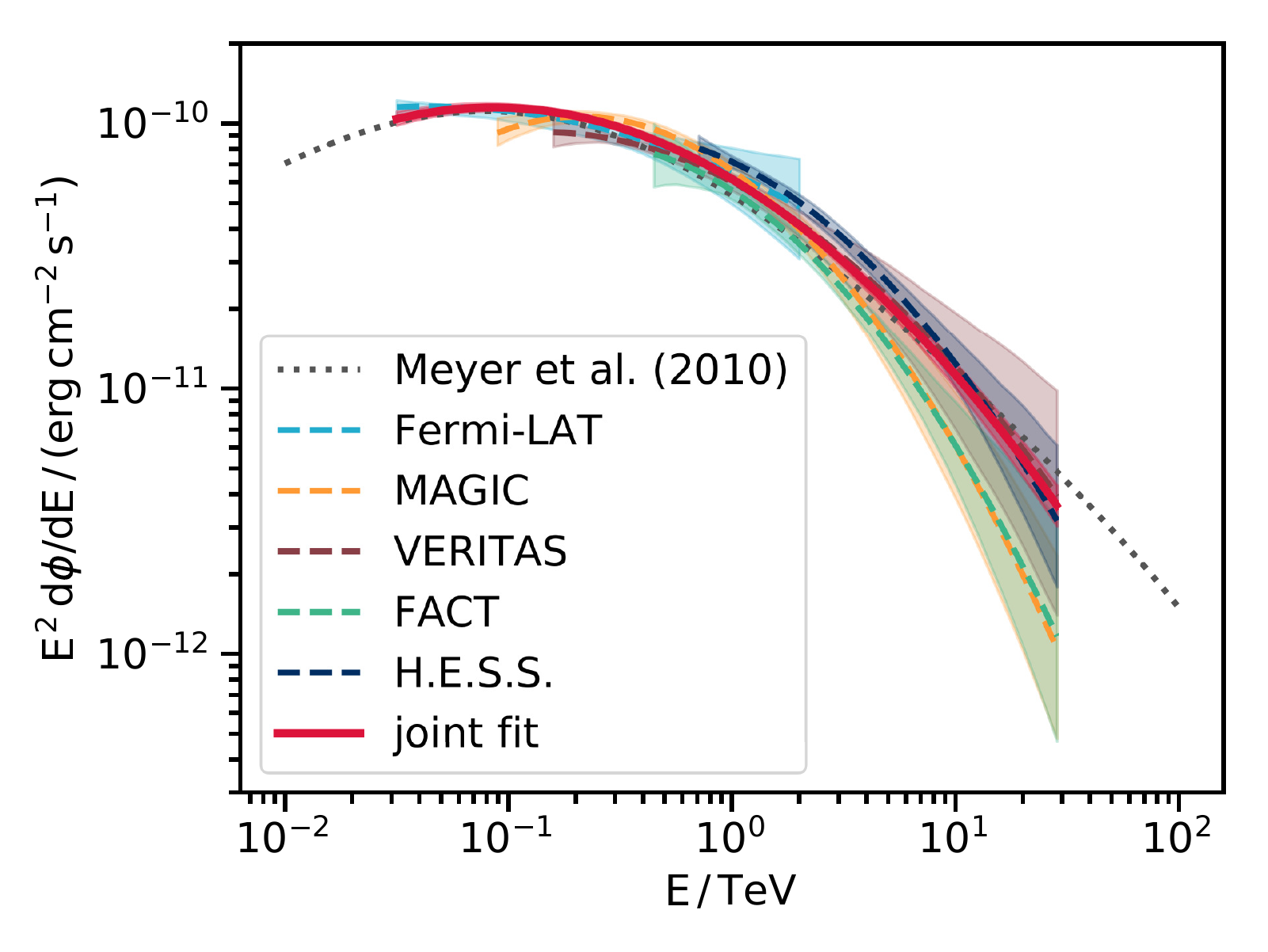}
    \caption{Fit of the Crab Nebula spectrum obtained using the individual datasets in \citet{joint_crab} and combining them in a single likelihood (red).}
    \label{fig:joint_crab_sed}
\end{figure}

The project, described in \citet{joint_crab}, provides the first example of fully-reproducible, multi-instrument analysis of gamma-ray data. Observations of the Crab Nebula by \texttt{Fermi}-LAT and by four of the operating IACTs (MAGIC, VERITAS, FACT and H.E.S.S.\footnote{The H.E.S.S. dataset is a taken from the H.E.S.S. public data release.}) converted into DL3 format, are analysed with \texttt{gammapy}. A one-dimensional binned likelihood analysis (obtaining photon counts versus energy via aperture photometry) is performed to estimate the spectrum of the Crab Nebula using all the datasets jointly (Figure \ref{fig:joint_crab_sed}). A possible modification of the likelihood to account for the systematic uncertainties in the energy scales of the different instruments is also discussed. The project constitute the first joint data release by IACTs. Full reproducibility is ensured by providing all the data, analysis scripts and tutorial notebooks publicly on \texttt{github}\footnote{\href{https://github.com/open-gamma-ray-astro/joint-crab}{https://github.com/open-gamma-ray-astro/joint-crab}} along with a \texttt{conda} environment installing all the software dependencies. An ancillary \texttt{docker} container\footnote{\href{https://hub.docker.com/r/gammapy/joint-crab}{https://hub.docker.com/r/gammapy/joint-crab}} guarantees a reproducibility longer than the life span of the software used.

\subsection{Analysis of the H.E.S.S. public data release with ctools}
\begin{figure}
    \centering
    \includegraphics[scale=0.23]{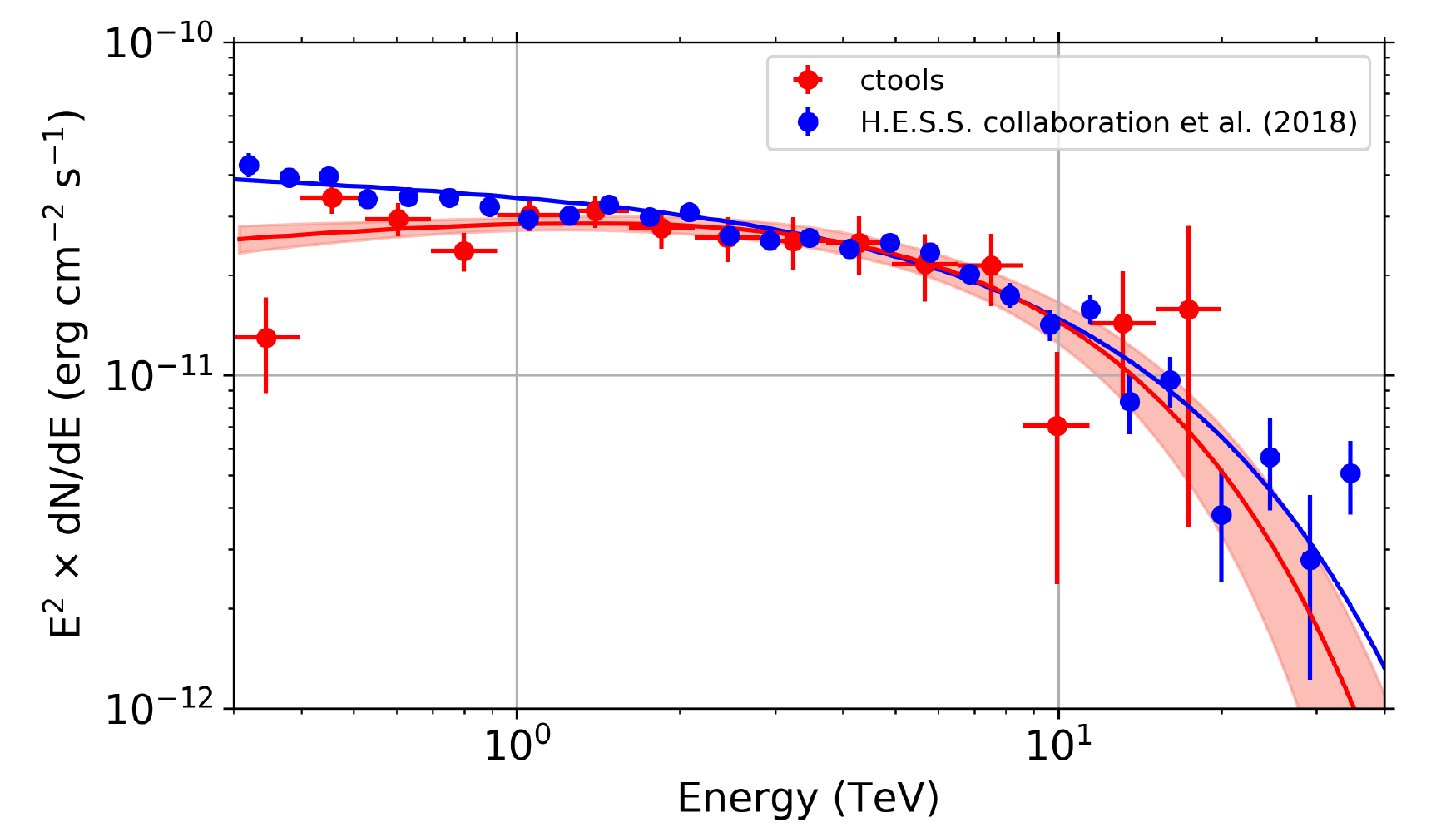}
    \includegraphics[scale=0.1]{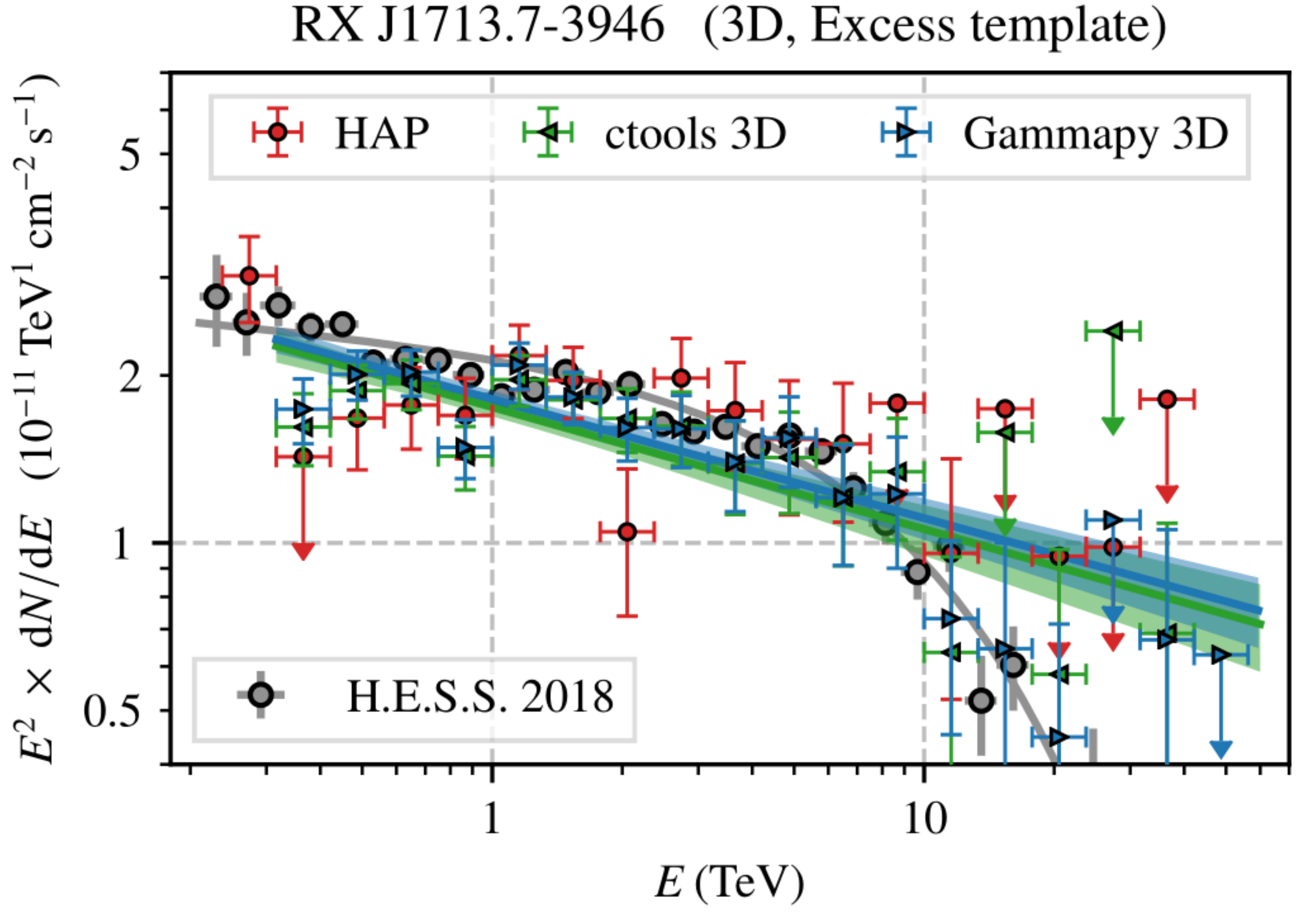}
    \caption{Validation of science tools on the extended source RX~J1713.7-3946. Left: comparison of the result of the \texttt{ctools} unbinned analysis and background model in \citet{ctools_paper} against the literature. Right: comparison of the \texttt{ctools} unbinned and \texttt{gammapy} binned analysis using the background model in \citet{validation_paper} against the H.E.S.S. propiretary software and the literature.}
    \label{fig:validation}
\end{figure}

The H.E.S.S. DL3 data release was analysed with \texttt{ctools} in \citet{ctools_paper}. The science tool was thus far tested only with CTA simulations and the publication represents the first test of its capabilities with actual IACT data. A recipe is presented to build a background model starting from the Off observations, the latter is used for spectral (and morphological - where possible) analysis of the 4 sources included in the H.E.S.S. DL3 data release. Results of binned and unbinned three-dimensional likelihoods (accounting at once for sources position, morphology and spectra) are compared against the ones of one-dimensional binned likelihoods. The internal consistency of the \texttt{ctools} results obtained with the different likelihood methods and their agreement with the results from literature extracted from larger datasets of the same sources show the maturity of the analysis tool (Figure \ref{fig:validation}). The paper also illustrates the capability of \texttt{ctools} to simultaneously analyse \textit{Fermi}-LAT and IACT data within the same likelihood framework.

\subsection{Validation of Open-source Science Tools and Background Model Construction in $\gamma$-ray Astronomy}
\citet{validation_paper} aim at validating both \texttt{ctools} and \texttt{gammapy} by comparing their results when analysing the H.E.S.S. DL3 data release. Here another procedure for the construction of a background model is illustrated, this time though relying on a large private dataset of H.E.S.S. off-source observations ($\sim 4000\,{\rm h}$). The results of three-dimensional and one-dimensional analyses by the two science tools are compared against the results obtained with one of the H.E.S.S. proprietary analysis chain on the same datasets and against the literature. For both likelihood methods an excellent agreement is found with the results of the H.E.S.S. analysis software.

\section{Conclusion}
The specifications of the \textit{Data Formats for Gamma-ray astronomy} forum have been recently employed to produce the first public IACT data release and to validate the analysis methods of the science tools currently in development. DL3 data constitutes the first example of multi-instrument gamma-ray data analysed with a fully-reproducible pipeline. The VHE community has been building crucial expertise in data standardisation that will be essential in the future operation of CTA.

\bibliography{B9-86}

\end{document}